\documentstyle[aps,preprint,tighten,graphicx,floats]{revtex}

\newcommand{\ben}{\begin{displaymath}}
\newcommand{\een}{\end{displaymath}}
\newcommand{\be}{\begin{equation}}
\newcommand{\ee}{\end{equation}}
\newcommand{\bea}{\begin{eqnarray}}
\newcommand{\eea}{\end{eqnarray}}
\newcommand{\eqn}[1]{\label{#1}}
\newcommand{\eq}[1]{Eq.~(\ref{#1})}

\begin{document}
\title{Doublet channel neutron-deuteron scattering in leading order
effective field theory}

\keywords{Document processing, Class file writing, \LaTeXe{}}

\author{B. Blankleider}
\address{SoCPES, The Flinders University of South Australia,
Bedford Park, SA 5042, Australia}
\author{J. Gegelia}
\address{INFN - Sezione di Ferrara, via Paradiso 12, 44100 Ferrara, Italy \\
and High Energy Physics Institute of TSU, University str. 9, Tbilisi 380086, 
Georgia}
\maketitle
\date{\today}

\begin{abstract}

The doublet channel neutron-deuteron scattering amplitude is calculated in
leading order effective field theory (EFT). It is shown that this amplitude does
not depend on a constant contact interaction three-body force. Satisfactory
agreement with available data is obtained when only two-body forces are
included.

\end{abstract}
\date{\today}

\section{Introduction}

An intriguing difficulty arises in the application of leading order EFT to the
three-body problem. One finds that the full amplitude describing three-boson
scattering, or nucleon-deuteron (nd) scattering in the $J=1/2$ channel, is
sensitive to the cutoff used to solve the scattering equations - even though
each perturbation diagram, with resummed two-body interactions, is individually
finite.  In \cite{Bedaque:1999kg} it was argued that the addition of a
one-parameter three-body force counter-term is necessary and sufficient to
eliminate this cutoff dependence. On the other hand, in refs.\
\cite{Gegelia:2000iz} and \cite{Blankleider:2000vi} we have shown, on the
example of three bosons, that the cutoff dependence is just a natural
consequence of the existence of infinitely many solutions to the given
scattering equation; moreover, by carefully identifying the physical amplitude
from amongst the infinitely many non-physical solutions, we have shown that the
cutoff problem can be solved without the introduction of a three-body force.

In the present contribution we show that the physical three-body
scattering amplitude in fact does not depend on the constant contact interaction
three-body force at all. Further, we demonstrate that for doublet channel nd
scattering, good agreement with experiment is obtained with the inclusion of
two-body forces only.


\section{Why there is no three-body force dependence}

To show that the leading order EFT three-body amplitude is independent of
constant three-body forces, it is sufficient to restrict the discussion to the
case of three bosons.

\subsection{Scattering amplitude without three-body forces}

In the three-boson case without three-body forces, the s-wave
particle-bound-state scattering amplitude $a(p,k)$ satisfies the equation
\cite{Bedaque:1999km}
\begin{equation}
\label{haeq}
a(p,k)
=M(p,k)+\frac{2\lambda}{\pi}\int_0^\infty dq\ M(p,q)
\frac{q^2}{q^2-k^2-i\epsilon} a(q,k),
\end{equation}
where
\begin{equation}
M(p,q)= \frac{8}{3}\left(\frac{1}{a_2}+\sqrt{\frac{3}{4}p^2-m E}\right)
   \left[ \frac{1}{2 pq}{\rm ln}
    \left(\frac{q^2+p q+p^2-m E}
               {q^2-q p+p^2-m E}\right)\right].  \eqn{M}
\end{equation}
\noindent
In this equation $k$ ($p$) is the incoming (outgoing) momentum magnitude, $E =
3k^2/4m - 1/ma_2^2$ is the total energy, and $a_2$ is the two-body scattering
length.  Here it is assumed that the summation of perturbation theory diagrams
and loop integration can be interchanged in the sense that the difference is of
higher order and hence negligable in given leading order calculations. In
general such assumptions have to be investigated very carefully as they may lead
to fictitious fundamental problems \cite{Gegelia:1998xr}.

Eq.\ (\ref{haeq}) is known as the S-TM equation \cite{skorny}, and in the
three-boson case has $\lambda=1$.  Three nucleons in the
spin $J=1/2$ channel obey a pair of integral equations with similar properties
to this bosonic equation, while the $J=3/2$ channel corresponds to
$\lambda=-1/2$. For $\lambda>0$ Danilov's work \cite{danilov} shows that
the homogeneous equation corresponding to \eq{haeq}
has a solution for arbitrary $E$; in particular, there exists a solution
for every energy corresponding to the {\em scattering} of
a projectile off a two-body bound state. 

The existence of these solutions implies that \eq{haeq} has an infinite number
of solutions. In fact the homogeneous equation has more than one solution for
any given $E$. Writing these solutions as $a_h^i$ where $i=1,2,3,\ldots$, the
most general solution of \eq{haeq} can be written as $a = a_{p} + \sum_i C_i
a_h^i$ where $a_{p}$ is any particular solution. It is useful to examine the
asymptotic behaviour of $a(p,k)$ for large $p$.  Because the inhomogeneous term
$M$ behaves asymptotically as $1/p$, it follows that either (i) $a\rightarrow 0$
faster than $1/p$, or (ii) the asymptotic behaviour of $a$ is determined by the
asymptotic behaviour of the homogeneous solution $a_h$. In the latter case the
asymptotic behaviour has the form \cite{danilov}
\be
a(p,k)= \sum_i A_i\left( k\right)p^{s_i}+O\left( {1/p}\right) \eqn{aasymptotic}
\ee
where $s_i$ are roots of the equation
\be
1-\frac{8\lambda}{\sqrt{3}}\frac{\sin\pi s/6}{s \cos\pi s/2}=0. \eqn{seq}
\ee
The summation in \eq{aasymptotic} goes over all solutions of Eq.\ (\ref{seq})
for which $|{\rm Re} s|<1$.  For $\lambda =1$ \eq{seq} has two roots for which
$|{\rm Re} s|<1$: $s=\pm is_0$, where $s_0\approx 1.00624$, so that
\eq{aasymptotic} gives the asymptotic behaviour of the amplitude as
\begin{equation}
a(p,k)\sim A_1\left( k\right)p^{is_0}+A_2\left( k\right)p^{-is_0}
\label{aasymptotics}.
\end{equation}
By contrast, for $\lambda <0$ the
homogeneous equation has no non-trivial solution and the
solution of \eq{haeq} is unique; in this case the physical amplitude $a$ must
vanish asymptotically faster than $1/p$.

In refs. \cite{Gegelia:2000iz,Blankleider:2000vi} we have shown that
the oscillatory behaviour of the general amplitude $a$ for $\lambda=1$ is simply
an artifact of the homogeneous equation (corresponding to \eq{haeq}) having
non-zero solutions, and that the {\em physical} amplitude does not display this
spurious behaviour; instead, it behaves just like the solution for $\lambda <0$,
namely, it vanishes asymptotically faster than $1/p$. Thus amongst the solutions
given by \eq{aasymptotics}, the physical solution is the one with
$A_1(k)=A_2(k)=0$. More generally, for any $\lambda > 0$ the physical amplitude
is the one that has no admixtures of homogeneous equation solutions, and by the
above argument, it must therefore vanish asymptotically faster than $1/p$.

Considering \eq{haeq} for the case where $a$ is the physical amplitude, since
the free term of \eq{haeq} behaves like ${1}/{p}$ for large $p$, the coefficient
of $1/p$ coming from this inhomogeneous term should cancel the coefficient of a
similar term coming from the integral part (we note that this argument is valid
for both $\lambda =1$ and $\lambda =-1/2$).  Hence
\be
\label{k+eqhoaskoeff}
0=\frac {4}{\sqrt{3}}+\frac {8\lambda}{\sqrt{3} \pi }\int_0^\infty  
\frac{dq \ q^2}{q^2-k^2-i\epsilon} \ a(q,k) .
\ee
\subsection{Identical scattering amplitude with a constant three-body force}
Multiplying \eq{k+eqhoaskoeff} by $2/\sqrt{3}(1/a_2+\sqrt{{3}/{4}p^2-m E})H$ and
adding the result to \eq{haeq}, we obtain the scattering equation where the
constant three-body force $H$ is included to all orders (an $H$ is simply added
to the term in the square bracket of \eq{M}). Hence the physical non-oscillating
solution of \eq{haeq} with no three-body force also satisfies the modified
\eq{haeq} where an arbitrary $H$ is included. Hence the inclusion of a
constant three-body force has no effect on the physical scattering amplitude.


In a recent paper \cite{Gabbiani:2001yh} it has been shown that doublet channel
neutron-deutron scattering amplitude in EFT with effective range parameters
taken into account, does not exhibit any dependence on a constant three-body
force. This result is therefore in agreement with the observations of the
present work.


\section{Nd scattering with two-body forces only}
Doublet channel neutron-deuteron scattering in leading order EFT is analogous
to the scalar case and does not involve any additional problems.  
%
Starting from the EFT Lagrangian, one can obtain the following system of doublet
channel neutron-deuteron scattering equations \cite{skorny,Bedaque:2000ve}:
\bea
a(p,k)&=&M(p,k)+\frac{2}{\pi}\int_0^\infty dq\ M(p,q)\left[ G_1(q)a(q,k)+
3 G_2(q)b(q,k)\right], \eqn{aeq}\\
b(p,k)&=&3 \ M(p,k)+\frac{2}{\pi}\int_0^\infty dq\ M(p,q)\left[ 3  G_1(q)a(q,k)+
G_2(q)b(q,k)\right] \eqn{beq},
\eea
where $a$ and $b$ are the neutron-$^3S_1$ (nd) and nucleon-$^1S_0$ amplitudes, 
respectively,
\[
G_1(q)=\frac{q^2}{q^2-k^2-i\epsilon}, \hspace{5mm}
G_2=\frac{{3}/{4} \ q^2}{\left( \sqrt{ {3}/{4}q^2-mE}+
{1}/{ a_2^t}\right)\left( \sqrt{{3}/{4}q^2-mE}-
{1}/{a_2^s}-i\epsilon \right)},
\]
$a_2^{s,t}$ are the two-particle scattering lengths in singlet and triplet
channels, and $E = 3k^2/4m - 1/m(a_2^t)^2$ is the total energy. Apart from a
factor of 1/4, the driving term $M(p,q)$ differs from \eq{M} only in that $a_2$
is replaced by $a_2^t$.
\begin{figure}[t]
\centerline{\includegraphics[width=8cm]{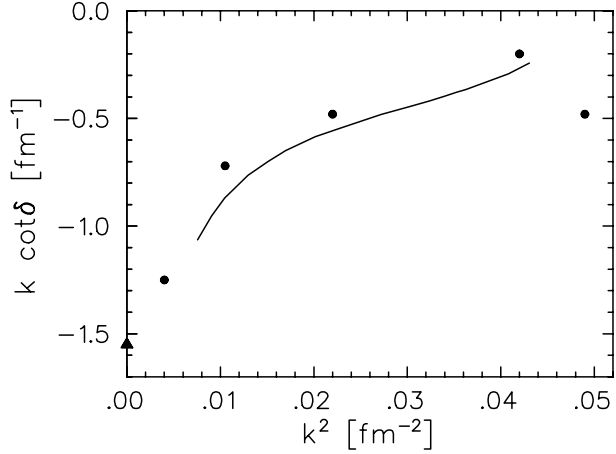}}
\vspace{2mm}

\caption{Doublet channel nd scattering phase shifts. Data are from 
phase shift analysis of van Oers and Seagrave 
(dots) and a measurement by Dilg et al. 
(triangle).}
\label{fig1}
\end{figure}

%
%
%
%

Solving the system (\ref{aeq})-(\ref{beq}) and isolating the non-oscillating
solution for $a(p,k)$ we obtain the physical amplitude for $nd$ scattering. The
results are shown in Fig.\ 1. In order to isolate the physical amplitude we had
to solve the corresponding homogeneous system of equations. This task is
especially difficult due to the existence of a continuum of solutions
corresponding to the continuous spectrum of scattering energies. The method
employed to achieve this numerical solution does not allow us to obtain high
accuracy for low momenta \cite{Gegelia:2000iz,Blankleider:2000vi} - that is
why our curve does not extend to the origin.  We note that while our
calculations fit the experimental data quite well, the accuracy of these data is
open to question. Ref.\ \cite{vanoers} does not contain error estimates and
ref.\ \cite{dilg} claims that at least the scattering length calculated in ref.\
\cite{vanoers} may be incorrect.
\vspace{10mm}


This work has been supported by a grant from the Australian Research Council. 


\end{document}